# Sub-Wavelength Multi-Channel Imaging Using a Solid Immersion Lens: Spectroscopy of Excitons in Single Quantum Dots

K. P. Hewaparakrama, S. Mackowski*, H. E. Jackson, and L. M. Smith

*Department of Physics University of Cincinnati, 45221-0011 Cincinnati OH, USA*

G. Karczewski and J. Kossut

*Institute of Physics Polish Academy of Sciences, 02-668 Warszawa, POLAND*

We demonstrate sub-wavelength imaging of excitons confined to single CdTe quantum dots. By combining slit-confocal microscopy with a hemispherical solid immersion lens we simultaneously map the emission of thousands of single quantum dots with a spatial resolution of 400 nm. By analyzing the linear polarization of the quantum dot emissions at B=0T, we find that the distribution of the exchange splitting is centered at zero with a standard deviation of ±340 µeV. Similar experiments performed at B=3T give an average value of the exciton effective g factor of 3.1±0.4. This experimental approach provides an effective means to gain statistical information about the quantum dot exciton fine structure in the ensemble.

*Author to whom the correspondence should be addressed: electronic mail: seb@physics.uc.edu

During the last decade, optical spectroscopy combined with sub-wavelength spatial resolution has enabled detailed probes of excitons confined to single quantum dots (QDs). One popular technique is to use a scanned [1] or fixed [2] sub-wavelength aperture, which is kept in the near-field with respect to the sample. Although a relatively high spatial resolution can be achieved (~100nm), a very low light transmission through the aperture limits the possibilities of optically mapping large sample areas. Another approach has utilized the more recent developments of the solid immersion lens (SIL) [3]. The collection efficiency when using a hemispherical SIL (h-SIL) can be enhanced by the index of refraction (n) [4-8]. At the same time, the spatial resolution is improved by the same factor [7].

In the previous experiments (using either scanned aperture or SIL) the emission of QDs has been imaged by collecting photoluminescence (PL) data for each pixel within the imaging region of the sample surface [5,9]. A tightly focused laser spot excited the QDs point-by-point in a raster-scanning mode and the emission has been detected by a CCD camera. Such an approach is extremely time-consuming, as imaging a 10 x 10 $\mu m^2$ surface with submicron resolution might typically require a total experimental time of tens of hours.

In this paper, we show how direct imaging of single QD emission can be dramatically improved using slit-confocal microscopy combined with a SIL. We utilize the fact that a two-dimensional pixel matrix of a CCD camera measures spectra for every point along the slit, and so all QDs within the slit can be imaged simultaneously in a single average. In this way, the experimental time for accumulating a macroscopic high-resolution image is reduced by over an order of magnitude. As a first demonstration of the utility of this technique, we show that we can uncover the fine structure of hundreds of single CdTe QDs through polarization-resolved imaging of the same ensemble of QDs with extremely high signal to noise ratio.

The QD sample was grown by molecular beam epitaxy on (100) oriented GaAs substrate. In order to form CdTe QDs, 4 monolayers of CdTe were deposited on a ZnTe buffer. Finally, the dot layer was capped by a 50 nm thick layer of ZnTe. Further details of the growth as well as the optical properties of this QD structure have been presented elsewhere by S. Mackowski [10].

A schematic diagram of the experimental setup used to image emission of individual CdTe QDs is displayed in Fig. 1(a). A 3 mm hemisphere SIL (h-SIL) (Schott LaSFN9 glass, n (600 nm) = 1.83) is fixed to the sample surface using a weak spring. The h-SIL increases the magnification and collection efficiency by factors of 2 and 4, respectively, over the bare objective [6]. The sample is mounted on the copper cold finger of a continuous-flow helium cryostat (temperature T=6 K), and external magnetic fields may be applied up to 4T. The polarization of the emitted PL is analyzed using a Babinet-Soleil compensator and Glan-Thomson linear polarizer.

The focus settings of the excitation and emission collection can be independently controlled. The excitation laser beam is expanded on the sample surface by a defocusing lens (see Fig. 1(a)) to a 30 µm spot so that a macroscopic region of the sample is excited with only a very weak power density (less than $1\mu W/1\mu m^2$). We estimate that on average only 1/3 of the dots are occupied with a single exciton at any time. The photoluminescence (PL) is collected in a back-scattering geometry using the h-SIL and a 50X/0.5NA long-working-distance microscope objective. In this way a 500X magnified image of the sample is focused onto the entrance slit of a DILOR triple spectrometer and detected by a 2000x800 pixel CCD detector. The entrance slit thus defines a 15 µm x 400 nm collection region so that the acquired CCD image (displayed in Fig. 1(b)) shows the position of the QD emitters along the slit direction (X) as well as its photon energy (E).

In order to acquire a complete two-dimensional (X-Y) image of the sample, a series of CCD images is recorded as the 400 nm x 15 µm image of the 200 µm x 7.5 mm slit is scanned 8 µm across the sample surface (Y direction in Fig. 1(c)) in a series of small steps (400 nm) using a piezoelectric translation stage. The spectrometer, CCD detector and translation stage are computer-controlled, and the series of two-dimensional CCD images are assembled into a 3D array. As a result, within 30 minutes, a complete image (8 x 15µm) of the sample surface is obtained with 400 nm spatial resolution at each of 2000 wavelengths, so that up to 60,000 CdTe QDs may be sampled in a single scan. It is important to note that the resolution along the X direction (along the slit) is determined by the optical resolution of the image and of the 3-axis DILOR spectrometer and the pixel size of the CCD camera. On the other hand, the resolution along the Y-direction is determined by the optical resolution of the image and the size of the entrance slit. Therefore, we have chosen a slit size of 200 µm in order to obtain similar spatial resolutions along X and Y directions (~400 nm).

In order to demonstrate capabilities of this PL imaging setup, we have performed polarization-resolved measurements of single CdTe QDs with and without an external magnetic field. In the first case (at B=0T), the fine structure of the excitons in single QDs is determined by the symmetry of the QD potential [11,12]. In cylindrically symmetric QDs, the emission is unpolarized since the optically active exciton states are the degenerate spin states $|+1\rangle$ and $|-1\rangle$. However, if the confinement potential of a QD is elongated along a particular direction, the eigenstates become the symmetric and antisymmetric linear combinations ($|x\rangle = (|+1\rangle + |-1\rangle)/\sqrt{2}$ and $|y\rangle = (|+1\rangle - |-1\rangle)/\sqrt{2}$) of the original spin states. These wave-functions have different energies, so that two observed emission lines are polarized parallel and perpendicular to the axis of the deformed potential [12]. In the second case (B>0T), an external magnetic field applied in the Faraday configuration leads to a Zeeman splitting of the QD exciton levels [12]. For symmetric QDs the Zeeman interaction lifts the degeneracy between $|\pm 1\rangle$ spin states; a single dot emission splits into two circularly ($\sigma+$ and $\sigma-$) polarized lines. On the other hand, for asymmetric QDs, an increase of the splitting is observed, and a gradual change of the optical selection rules leads eventually to circularly polarized emission [12]. We note that the observation of polarized emission lines requires that the exciton spin relaxation time be longer than the lifetime as indeed has been observed for these CdTe QDs [13].

The exchange splitting of the exciton states in single CdTe QDs has been analyzed by measuring two complete three-dimensional (X-Y-Energy) sets of images for two orthogonal linear polarizations of the analyzer, $\pi_x$ and $\pi_y$. In Fig. 2(a) we show a subset of the complete $\pi_x$ (gray-scale shaded contour map) and $\pi_y$ (black line contour map) polarized images of QD emission as two contour maps. It is important to note that in these two scans, taken 30 minutes apart, the same region of the sample was imaged and, therefore, a large number of symmetric and asymmetric QDs can be directly measured and the exchange splitting determined. The maps are obtained at photon energies of 2.01765 eV and 2.01776 eV for $\pi_x$ and $\pi_y$ linear polarizations, respectively. As can be seen in this plot, the positions of most of the small circular high intensity regions on the gray scale and black line contour maps are overlapping. This implies that these high intensity small regions are images of single QD emissions. Two characteristic QDs, which are discussed in detail below, are marked QD1 and QD2 in this plot.

Figure 2 (b) and (c) shows the linearly polarized PL spectra taken at the center of the dots marked QD1 and QD2 in Fig. 2 (a). As can be seen in Fig 2 (b), the exciton emission from QD1 consists of a doublet which is linearly polarized along $\pi_x$ and $\pi_y$, indicating that the dot is asymmetric with an exchange energy splitting of 120 $\mu$eV. On the other hand, QD2 (see Fig.

2(c)) is approximately symmetric since the emission shows only a single line, which is unpolarized. We have analyzed over 80 dots and, as shown in Fig. 2 (d), the exchange splitting has a normal distribution around zero, with a standard deviation of ±340 μeV.

We study the Zeeman splitting of excitons in the CdTe QDs by using magneto-PL imaging. To determine the g factor, we directly image the same region of the sample at 3T for the two circular polarizations (σ+ and σ−). Figure 3 (a) shows a subset of the complete σ+ and σ- polarized QD images. The gray-scale shaded contour map and the black line contour map are the maps obtained for energies of 2.0481 eV and 2.0476 eV, for σ+ and σ- polarized QD images, respectively. As can be seen, the high intensity region of the gray shaded image and the center of the black line image overlap. This indicates that these are two polarized images of the emission of the same QD. Fig. 3 (b) shows the circularly polarized spectra of this QD. The energy splitting of these peaks is 510 μeV, which corresponds to an exciton g factor of 2.9. Again, from these two 30 minute scans, we have analyzed more than 80 dots and, as can be seen in Fig. 3 (c), the effective exciton g-factors for these CdTe dots are normally distributed with a value of 3.1±0.4.

We want to point out that these experiments were accomplished with only modest spatial resolution, which could be further enhanced by using either a higher index SIL material, like $ZrO_2$ (n=2.16) [4] or GaP (n = 3.3) [5]) or a (truncated) Weierstrass SIL [8]. Moreover, this imaging technique may easily be applied to probe nearly all two-dimensional arrays of nanoscale structures.

In summary, by combining slit-confocal microscopy with hemisphere solid immersion lens we are able to directly image the emission of large number of CdTe QDs. The high collection efficiency of the SIL, combined with parallel data acquisition along the spectrometer slit, reduces by orders of magnitude the acquisition time compared to previous raster methods. As an example of the power of this method we have used polarization-resolved imaging of single QDs to determine the exciton fine structure of a large distribution of QDs in a single experiment.

We acknowledge the support of NSF through Grants DMR 0071797 and 0216374 (United States) and PBZ-KBN-044/P03/2001 (Poland).

**References**

[1] H.F. Hess, E. Betzig, T.D. Harris, L.N. Pfeiffer, and K.W. West, Science **264**, 1740 (1994).

[2] D. Gammon, E.S. Shaw, B.V. Shanabrook, D.S. Katzer, and D. Park, Science **273**, 87 (1996).

[3] S. M. Mansfield and G. S. Kino, Phys. Rev. **B 57**, 2615 (1990).

[4] S. Moehl, Hui Zhao, B. Dal Don, S. Wachter, and H. Kalt, J. Appl. Phys. **93**, 6265 (2003).

[5] Q. Wu, R. D. Grober, D. Gammon, and D. S. Katzer, Phys. Rev. Lett. **83**, 2652 (1999).

[6] V. Zwiller and G. Bjork, J. Appl. Phys. **92**, 660 (2002).

[7] B. B. Goldberg, S. B. Ippolito, L. Novotny, L. Zhiheng, and M. S. Unlu, IEEE J. Selected Topics in Quantum Electronics **8**, 1051 (2002).

[8] K. Karrai, X. Lorenz, and L. Novotny, Appl. Phys. Lett. **77**, 3459 (2000).

[9] K. Matsuda, T. Saiki, S. Nomura, M. Mihara, Y. Aoyagi, S. Nair, and T. Takagahara Phys. Rev. Lett. **91**, 177401 (2003).

[10] S. Mackowski, Thin Solid Films, **412**, 96 (2002).

[11] M. Bayer, A. Kuther, A. Forchel, A. Gorbunov, V. B. Timofeev, F. Schäfer, and J. P. Reithmaier, Phys. Rev. Lett. **82**, 1748 (1999).

[12] V.D. Kulakovskii, G. Bacher, R. Weigand, T. Kummel, A. Forchel, E. Borovitskaya, K. Leonardi, D. Hommel, Phys. Rev. Lett. **82**, 1780 (1999).

[13] S. Mackowski, T. A. Nguyen, H. E. Jackson, L. M. Smith, J. Kossut, and G. Karczewski, Appl. Phys. Lett. **83**, 5524 (2003).

**Figure captions**

Figure 1. (a) Experimental setup that combines slit-confocal microscopy with solid immersion lens. (b) The laser spot is defocused, which allow to directly image a 400 nm wide stripe of the sample to the spectrometer slit and the CCD matrix. (c) Then, by scanning the sample across the slit, a full three-dimensional X-Y-Energy dataset is collected.

Figure 2. (a) Two X-Y maps obtained for $\pi_X$ (gray scale) and $\pi_Y$ (black line) polarizations at 2.01765 eV and 2.01776 eV, respectively. The measurement was performed at T=6K and at B=0T. (b) PL emission spectra of a single CdTe QD (QD1) extracted for both linear polarizations. (c) As in Fig. 1b, but for the QD2. (d) Histogram of the exchange energy splitting measured of 80 single QDs. The solid line is a Gaussian fit to the data.

Figure 3. (a) The spatial X-Y maps of the same single QD measured at B=3T in both $\sigma+$ and $\sigma-$ circular polarizations at energies of 2.0481 eV and 2.0476 eV, respectively. (b) Low temperature (T=6K) PL spectra of a single CdTe QD measured for both circular polarizations. These spectra have been extracted from the full three-dimensional datasets. (c) Histogram of the effective exciton g-factor distribution obtained of 80 single QDs.

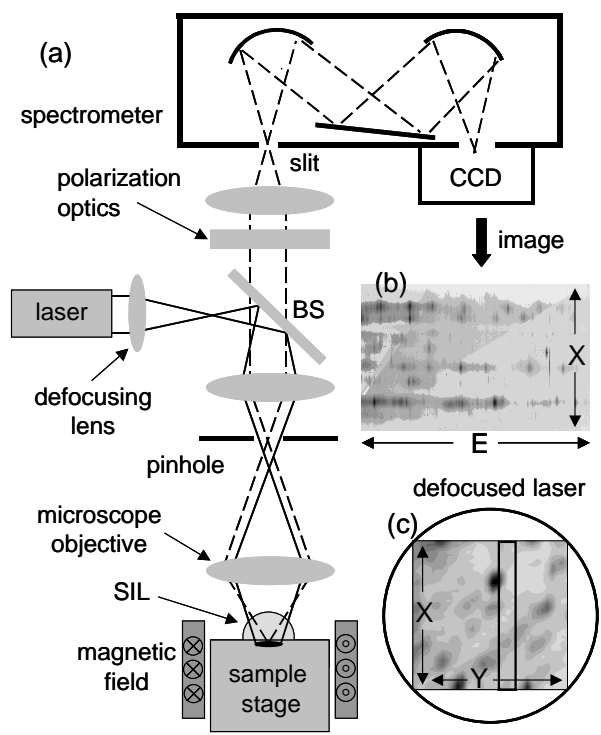

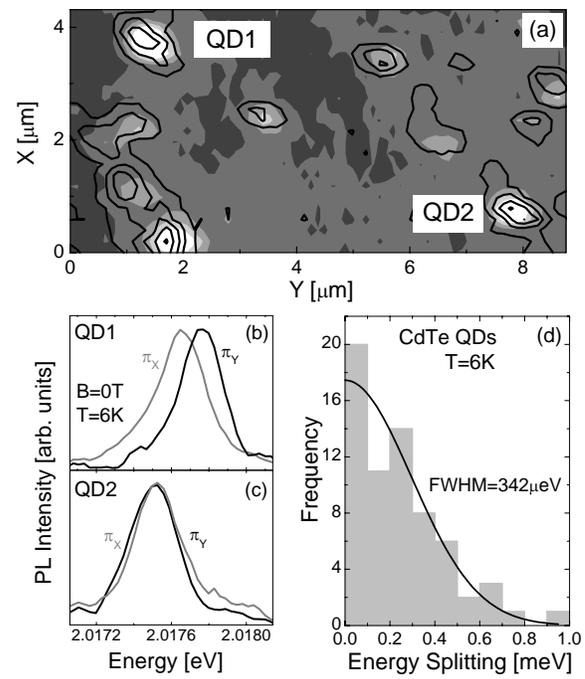

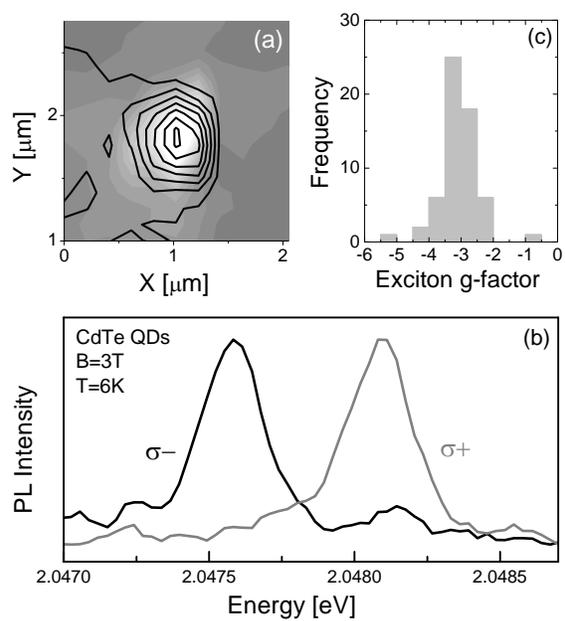